\documentclass[journal]{IEEEtran}
\usepackage{amsmath,graphicx,dsfont,ctable} 
\usepackage{algorithm}
\usepackage{algorithmic}
\usepackage{dblfloatfix}
\usepackage{amsfonts}
\usepackage{caption}
\usepackage{hyperref}
\usepackage{makecell}
\usepackage{perpage} 
\MakePerPage{footnote} 
\begin{document}
\title{Plane-Wave Ultrasound Beamforming Through\\
	Independent Component Analysis}
\author{Sobhan Goudarzi,
        Amir Asif
        \IEEEmembership{Senior Member,~IEEE,}
        and~Hassan Rivaz
        \IEEEmembership{Senior Member,~IEEE}
}
\maketitle
\begin{abstract}
Beamforming in plane-wave imaging (PWI) is an essential step in creating images with optimal quality. Adaptive methods estimate the apodization weights from echo traces acquired by several transducer elements. Herein, we formulate plane-wave beamforming as a blind source separation problem. The output of each transducer element is considered as a non-independent observation of the field. As such, beamforming can be formulated as the estimation of an independent component out of the observations. We then adapt the independent component analysis (ICA) algorithm to solve this problem and reconstruct the final image. The proposed method is evaluated on a set of simulation, real phantom, and \textit{in vivo} data available from the PWI challenge in medical ultrasound. The performance of the proposed beamforming approach is also evaluated in different imaging settings. The proposed algorithm improves lateral resolution by as much as $36.5\%$ and contrast by $10\%$ as compared to the classical delay and sum. Moreover, results are compared with other well-known adaptive methods. Finally, the robustness of the proposed method to noise is investigated.        
\end{abstract}
%
\label{sec:sec1}
\section{Introduction}
\IEEEPARstart{U}{ltrasound} imaging experienced a revolution with the introduction of Plane-Wave Imaging (PWI) in which frame-rate can reach several thousands per second. In contrast to other techniques, PWI fires all elements of the probe simultaneously to form a flat wavefront and span the whole region of interest in a single shot. This technique has been successfully applied to different applications such as imaging of shear waves, contrast imaging, and Doppler imaging of blood flow~\cite{6689779}. Having a non-focused transmitted beam, however, leads to poor resolution and low contrast in PWI. This drawback was addressed by coherent compounding of images obtained by several insonifications of different angles~\cite{4816058}. Consequently, there is always a trade-off between image quality and frame-rate. Hence, beamforming is witnessing a growing attention in order to enhance the quality of images without sacrificing the frame-rate.\par
In PWI, beamforming mainly refers to the method of merging the outputs of different crystal elements. More specifically, it is a weighting function across the probe which is referred  to as apodization. It can also be used during transmission. Delay-and-sum (DAS) is a classical nonadaptive beamforming method in which apodization weights for different pixels of the image are assigned based on the \textit{F}-number as well as a predefined window shape. As known from spectral estimation, there is often a trade-off between the width of main lobe and energy of side lobes of the apodization window. When measured backscattered signals are directly used to optimize the apodization weights, the beamforming method is considered adaptive.\par
Capon or minimum variance (MV) is a well-known adaptive method in which apodization weights are estimated to minimize the variance of output while preserving the unity gain in the steering direction~\cite{5278437}. Asl and Mahloojifar~\cite{5611687} extended eigenspace-based MV (EMV) to better suppress off-axis signals. The main issue with MV is the estimation of covariance matrix of data, which is time consuming and limits its real-time applications~\cite{4291510,6217562}.\par
Matrone \textit{et al.}~\cite{6960091} proposed Filtered-delay multiply and sum (F-DMAS). The pipeline of algorithm contains the multiplication of delayed radio-frequency (RF) signals followed by summation of signed square roots. Finally, the beamformed signal is achieved by bandpass filtering to remove the DC component.\par
A family of adaptive beamforming algorithms are based on phase coherence. First, coherence factor (CF) is defined as the ratio between the coherent and incoherent energy across the aperture~\cite{410562} and then used as an adaptive weight following the DAS beamformer to enhance the image quality~\cite{1182117}. CF was generalized to be computed from Fourier spectra over the aperture of the delayed channel data and in a range of low spatial frequency region~\cite{1182117}. Subsequently, Camacho \textit{et al.}~\cite{4976281} used phase information of aperture data to compute the adaptive correction weight and proposed phased CF (PCF). However, the estimated correction weights of CF methods can be affected by speckle noise.\par 
More recently and more specifically for PWI, the MV approach was applied in~\cite{2254293,7728906}. Nguyen and Prager~\cite{8259306} proposed extensions to MV for coherent plane-wave compounding (CPWC). Beamforming based on compressive sensing for PWI was introduced in~\cite{7329447,david2015time,7582552,8091286}. Dei \textit{et al.}~\cite{2255526,027001} investigated the performance of their beamforming method entitled aperture domain model image reconstruction (ADMIRE) on PWI. Beamforming in Fourier domain on PWI was introduced in~\cite{8359331}. This was further developed to incorporate coherent compounding and angular weighting in~\cite{8306445}. Beamforming as a regularized inverse problem was introduced in~\cite{7565515} and applied at each depth separately. This point of view was extended in~\cite{8052532} to solve inverse problem for all image
depths jointly. Recently, a statistical interpretation of beamforming entitle iterative maximum-a-posteriori (iMAP) was introduced in~\cite{8789467}.\par
Herein, we propose a new framework for adaptive plane-wave beamforming wherein apodization weights are estimated through independent component analysis (ICA). In the field of US imaging, ICA has been mainly used for clutter filtering and noise suppression~\cite{9005396,5456258,tierney2019independent,gallippi2003bss,4559657,gallippi2002adaptive}. Recently, ICA was used as a dimensionality reduction technique to speed up ADMIRE beamforming~\cite{8746209}.\par 
What motivates us to make use of ICA is a principal physical limitation which is brought about by wave propagation. More specifically, the backscattered waves from sources at the same time, leading to a single sensory data at the resulting RF signal. Fortunately, this distance is not the same for other elements and, therefore, a group of distinct sources are indistinguishable from the output of each piezoelectric element. Our approach considers the output of all piezoelectric elements as a set of non-independent observations of the field and then uses beamforming to extract independent components. As such, we reconstruct the whole image simultaneously. The performance of the proposed adaptive plane-wave beamforming is evaluated and compared to other methods on a set of simulation, phantom, and \textit{in vivo} data provided by PWI challenge in medical ultrasound (PICMUS) 2016~\cite{7728908}. Moreover, we run the algorithm on another set of publicly available phantom data recorded with an Alpinion ultrasound machine (Bothell, WA) to test its performance on images recorded with different settings. Finally, the robustness of proposed approach is compared to other methods in the presence of noisy observation.   
\section{METHODS}
\label{sec:sec2}
\begin{figure}[t!]
	\centering
	\centerline{\includegraphics[width=7.5cm]{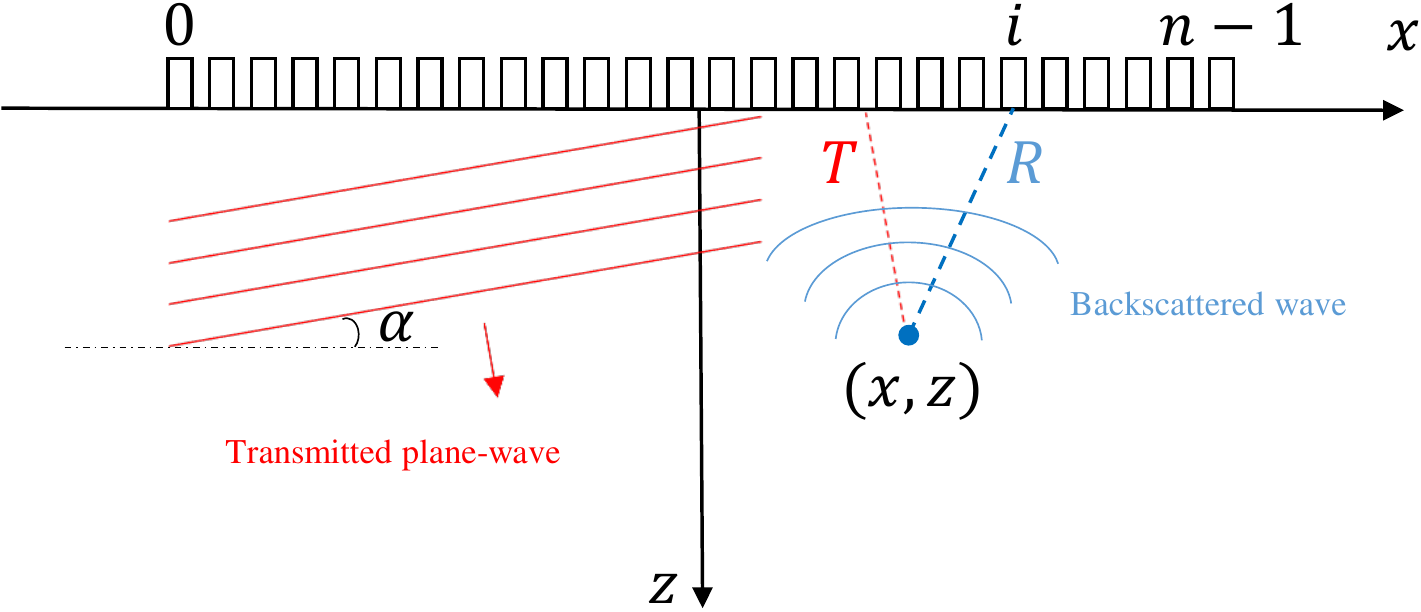}}
	\caption{Geometrical illustration of the PWI. Notation $\alpha$ is the steering angle of the incidence wave.}
	\label{fig:fig1}
\end{figure}
Consider a linear array of $n$ elements, symmetrically distributed on the $x$-axis, transmitting along the positive $z$-axis (Fig.~\ref{fig:fig1}). Let us assume a plane-wave with angle $\alpha$ spans the domain with a sound speed of $c$. The backscattered signals received by element $i$ located at $x_i$ is denoted by $\mathbf{h}_i(t)$. Without any loss of generality, $zcos(\alpha)+xsin(\alpha)$ is the transmission distance $d_t$ from the origin of the transmitted plane-wave to an arbitrary point $(x,z)$ in the region-of-interest (ROI) and $\sqrt{ (x-x_i)^{2}+z^2 }$ is the receiving distance $d_r$ from $(x,z)$ to the location of element $i$. The RF data corresponding to $(x,z)$ in $\mathbf{h}_i(t)$ can be found by applying the associated propagation delay as follows (hereafter, capital and bold font variables represent matrices and vectors, respectively):
\begin{equation} 
	\label{eq:1}
	\tau =\frac{d_t+d_r}{c}  \Longrightarrow  R_i(x,z) = \mathbf{h}_i(t)\mid_{t=\tau} ,
\end{equation}
where $R$ is the matrix containing RF data received by the entire transducer array. Each point $(x,z)$ of the final image can be obtained through the following weighted summation:
\begin{equation} 
\label{eq:2}
R(x,z) = \sum_{i=1}^{n} \mathbf{w}(i)R_i(x,z),
\end{equation}
where $\mathbf{w}$ is the apodization window of length $n$. In practice, however, we utilize dynamic beamforming where the \textit{F}-number (F) is fixed for the entire image. Therefore, the number of elements considered for the reconstruction of each depth of the image $l$ is calculated as follows:
\begin{equation} 
\label{eq:3}
F = z/l.
\end{equation}
Our goal is to estimate the apodization window using ICA.  
\subsection{Independent Component Analysis}
\label{sec:sec21}
ICA is a framework used to separate signal components mixed in observations recorded at different transducer elements~\cite{2000411}. Assuming an $n$-dimensional signal space, i.e., an $n$-dimensional observed data $\mathbf{x}$, $n$-dimensional independent sources $\mathbf{s}$, and a square transformation matrix $W$ of size $n\times n$, the mixing model can be written as follows~\cite{2000411}:
\begin{equation} 
\label{eq:6}
\mathbf{s}=W\mathbf{x}.
\end{equation} 
With the assumption of having independent and non-Gaussian sources (at the most one independent Gaussian source is allowed), both of $W$ and $\mathbf{s}$ can be estimated using the ICA algorithm. In practice, the objective function for ICA estimation can be formulated using different measures of non-Gaussianity such as kurtosis, negentropy, and mutual information. Moreover, it is very useful to center and whiten the observations before applying ICA. One of the most famous algorithms of ICA implementation is FastICA, where a unit vector $w$ is computed such that the dot product $\mathbf{w}^T\mathbf{x}$ maximizes negentropy. FastICA algorithm can be summerized in four steps as follows~\cite{2000411}:
\begin{enumerate}
\item Random initialization of vector $\mathbf{w}$.
\item $\mathbf{w}_{new} = E\{\mathbf{x}g(\mathbf{w}^T\mathbf{x})\}-E\{g'(\mathbf{w}^T\mathbf{x})\}\mathbf{w}$
\item $\mathbf{w} = \mathbf{w}_{new}/ \|\mathbf{w}_{new}\|$
\item Return to step $2$ until the direction of $\mathbf{w}$ does not change.
\end{enumerate}
where $E$ denotes the expectation operation. $g$ and $g'$ are first and second derivatives of a nonquadratic nonlinear function $f$, respectively. It was shown that either of the two functions $f$ is robust for negentropy estimation~\cite{hyvarinen1998new}:\\
$f(u)=\frac{1}{a_1}\log \cosh a_1u$ or $f(u)=-\exp(-u^2/2)$\\
where $1\leq a_1\leq 2$.\par
As discussed in~\cite{2000411}, ICA can be considered as a variant of the projection pursuit algorithm~\cite{1672644}, which enables one-by-one estimation of the independent components. This is an important feature since the computational load of the method substantially decreases in our case wherein only
one of the independent components is required. The aforementioned algorithm can be easily extended to estimate several independent components. More details regarding the FastICA algorithm can be found in~\cite{2000411}.
\subsection{Beamforming Using ICA}
\label{sec:sec22}
\begin{figure}[t!]
	\centering
	\centerline{\includegraphics[width=8.5cm]{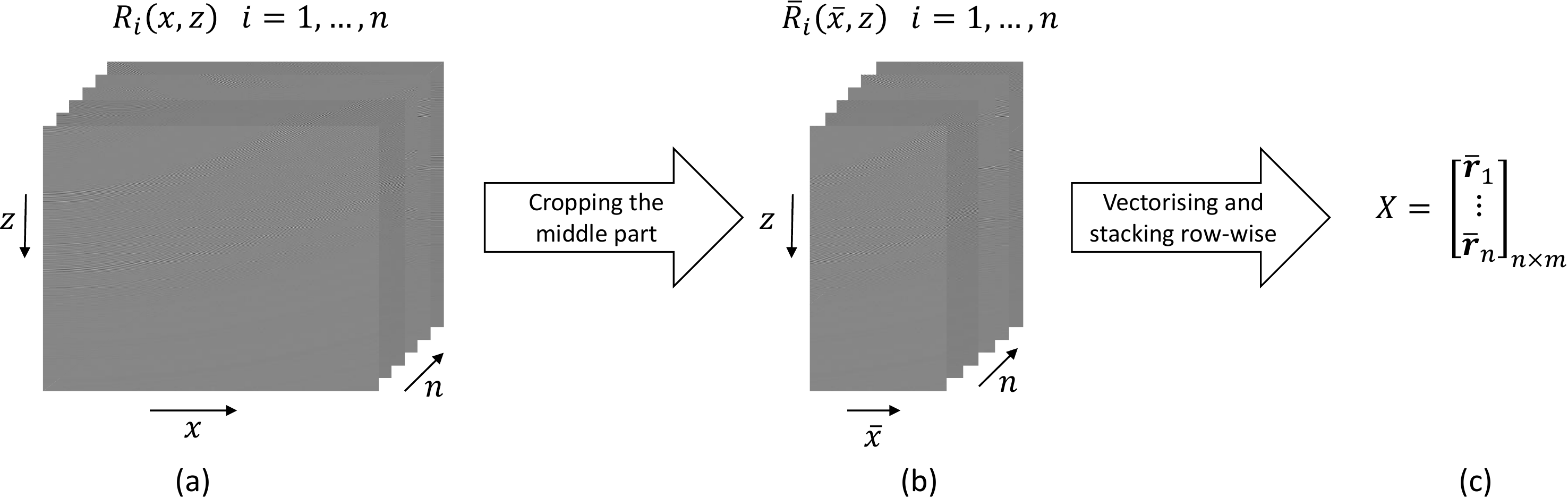}}
	\caption{Steps 1 and 2 of the proposed ICA beamforming. (a) From all crystal elements, RF data of each pixel ($R_i(x,z)$) is extracted by applying the propagation delays to received backscattered signals ($\mathbf{h}_i(t)$). (b) Central pixels of the image around which the crystals are symmetric are cropped from each $R_i(x,z)$. (c) The matrix of observation $X$ is constructed by stacking $\bar{\mathbf{r}_i}$.}
	\label{fig:fig9}
\end{figure}
In general, our goal is to reconstruct a high-quality spatial map of target echogenicity. More specifically, what we do is discretizing the map of scatterers that leads to pixels. Each pixel corresponds to an averaged tissue reflectivity function over the extent of the pixel. According to the central limit theorem, such a pixel has a Gaussian distribution if the number of diffusion scatters in the medium is significantly high~\cite{80001862243}. Therefore, ICA may appear not suited because it finds independent components that maximize non-Gaussianity. However, it is shown in~\cite{2000411} that the ICA can still be used even if only one of the independent components is Gaussian. Therefore, we look for the discretized tissue reflectivity function (i.e. the map of scatterers) as the only independent Gaussian source.\par
When the RF data corresponding to each pixel of the final image is extracted from the output of each element, there is also the possibility of source ambiguity. More specifically, the backscattered waves of at least two different scatterers arrive simultaneously. Without loss of generality, when $\alpha=0$, the backscattered waves of two distinct scatterers (with indices 1 and 2) arrive at the same time in element $i$ if and only if they have the same propagation delay $\tau$. Form Eq.~(\ref{eq:1}) and if the first scatterer is in the lateral position $x_i$, it can be written that:
\begin{equation} 
\label{eq:4}
\tau_1 = \tau_2 \Longrightarrow 2z_1=z_2+\sqrt{(x_2-x_i)^{2}+z_2^2}.
\end{equation}  
Simplifying (\ref{eq:4}) gives:
\begin{equation} 
\label{eq:5}
z_2=z_1-\frac{(x_2-x_i)^2}{4z_1}.
\end{equation}  
So, for $z_2<z_1$, there are a bunch of scatterers located on an ellipse, whose reflections arrive at the same time with scatterer $1$. In the continuous case, this problem is fully addressable. In the discrete case, however, there is the error of quantization as well. Although this problem was shown for the specific case of $\alpha=0$, the concept can be extended for different angles.\par
As can be seen from Eq.~(\ref{eq:5}), the group of scatterers from whom their reflections arrive simultaneously are not the same for each crystal element. This point provides the opportunity of source separation. Herein, the task is to extract an independent RF signal out of a set of non-independent observations. In the ideal case, the desired independent RF signal contains the trace of only one distinct scatterer in each sample. It has to be mentioned that in practice, the axial and lateral resolution are based on the sampling frequency of the system, the center frequency of the transmitted wave, and the transducer design. So, when we refer to one scatterer, it means one pixel of the final image.\par
In ultrasound beamforming, an issue is that the apodization window is not fixed throughout the image. More specifically, ICA works with a fixed transformation matrix $W$ in Eq.~(\ref{eq:6}). In ultrasound images, however, the apodization weight is not spatially invariant, rendering a different set of values of $W$ for different pixels. Two points make the apodization weight spatially variant. First, for pixels lying at the two lateral ends of the image, there are crystal elements predominately lying along one side. Second, as explained in section ~\ref{sec:sec2}, pixels of different depths of the image are reconstructed using a different number of elements to hold \textit{F}-number fixed in the entire image. Hence, if we do not consider these points, ICA fails to estimate the source and apodization windows, leading to images that are even lower in quality than DAS.\par
To solve the aforementioned problem, we consider only the central pixels of the image around which the crystals are symmetric as the input to the ICA algorithm. In this way, the cropped portion of $R_i$~(Eq.~\ref{eq:1}) is used to construct the observation matrix $X$. In our ICA formulation, therefore, the observations are RF data corresponding to central pixels of final image that are recorded by all crystal elements of the probe.\par
Our proposed beamforming algorithm for PWI using ICA includes the following steps:
\begin{enumerate}
	\item The propagation delay is applied to all channels of data in order to generate $R_i, i=0, ..., n-1$.
	\item Each $R_i$ is considered as an observation of the field (Fig.~\ref{fig:fig9},). First, it is cropped (denoted by $\bar{R}_i(\bar{x},z)$) and then vectorized in a row vector ($\bar{\mathbf{r}_i}$). Consequently, the observation matrix $X$ is constructed by stacking the row vectors ($\bar{\mathbf{r}_i}$). Finally, matrix $X$ is centralized and whitened before running ICA.  
	\item The independent source and corresponding mixing vector are estimated using the FastICA algorithm~\cite{761722} by maximizing Negentropy as the measure of non-Gaussianity. 
	\item The apodization window (the estimated transformation matrix in the last step) is applied throughout the image based on a predefined \textit{F}-number.  
\end{enumerate}
\begin{figure*}[b!]
	\centering
	\centerline{\includegraphics[width=\textwidth]{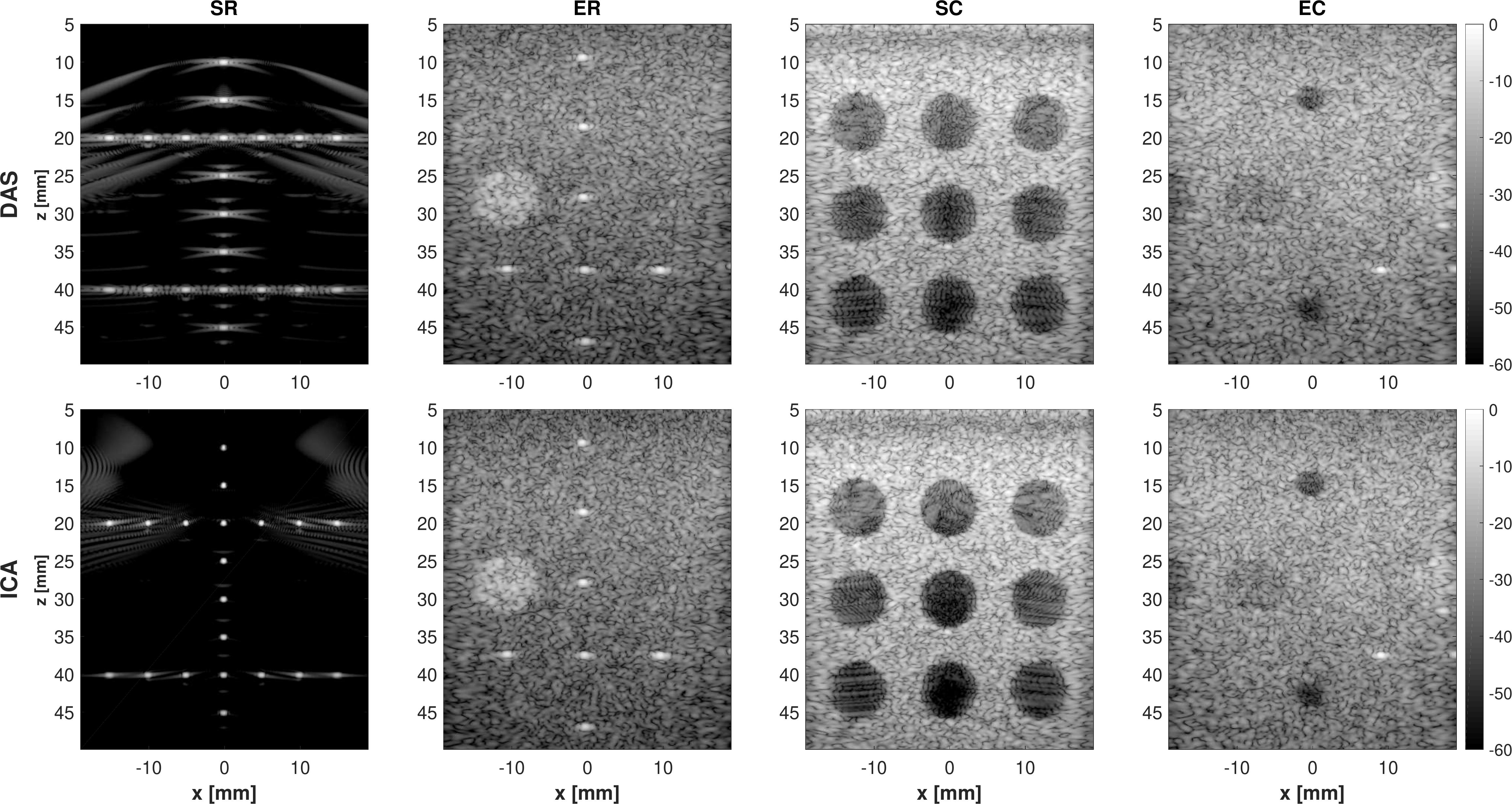}}
	\caption{Beamforming results on the single $0^{\circ}$ plane wave. Columns indicate different image data sets while rows correspond to beamforming methods.}
	\label{fig:fig2}
\end{figure*}
\section{EXPERIMENTS}
\label{sec:sec3}
\subsection{Dataset}
\label{sec:sec31}
Herein, we use a publicly available PWI dataset, entitled PICMUS, which was provided by the IEEE International Ultrasonics Symposiun (IUS 2016) in order to benchmark novel beamforming methods~\cite{7728908}. The PICMUS data utilized in this work include:
\begin{enumerate}
	\item Simulation resolution (SR): a simulated ultrasound image containing point targets distributed
	vertically and horizontally over an anechoic background designed to assess the performance of beamforming methods in terms of spatial resolution.
	\item Simulation contrast (SC): a simulated ultrasound image containing anechoic cysts distributed vertically and horizontally 
	over fully developed speckle designed to assess the performance of beamforming methods in terms of contrast.
	\item Experimental Resolution (ER): an experimental ultrasound image was recorded on a CIRS Multi-Purpose Ultrasound Phantom (Model 040GSE) in the regions containing several wires against speckle background to assess the performance of beamforming methods in terms of spatial resolution.
	\item Experimental contrast (EC): an experimental ultrasound image was recorded on the same phantom as ER but in the regions containing two anechoic
	cysts against speckle background to assess the performance of beamforming methods in terms of contrast.
\end{enumerate}
In addition, PICMUS dataset also contains two \textit{in vivo} ultrasound images, showing cross-sectional and
longitudinal views, recorded on the carotid artery of a volunteer. All of the phantom and \textit{in vivo} data were collected using a Verasonics Vantage 256 research scanner and a L11 probe (Verasonics Inc., Redmond, WA). The simulation settings were selected to be as similar as possible to the experimental setup.\par
For each of mentioned groups, a collection of received prebeamformed data corresponding to 75 steered Plane-Waves covering the angle span from $-16^{\circ}$ to $16^{\circ}$  was provided. Both RF and IQ (phase quadrature) formats of data were provided. The proposed algorithm works on the RF version of data. More details regarding PICMUS dataset can be found in~\cite{7728908}.\par
Another publicly available dataset on PWI is used to investigate the performance of the proposed method on data collected with other imaging settings. In this database, two datasets of CPWC recorded using an Alpinion scanner with a L3-8 probe from a CIRS phantom are used. These data sets are available through the ultrasound toolbox~\cite{8092389}, a MATLAB toolbox for processing ultrasonic signals. The first dataset was recorded from regions containing hyperechoic cyst and points scatterers, and the second one recorded from regions containing hypoechoic cyst. Imaging setting such as probe geometry, transmit frequency, sampling frequency, and etc. are completely different from PICMUS. More details regarding this data can be found in~\url{https://www.ustb.no/}.    
\subsection{Evaluation Metrics}
\label{sec:sec32}
Contrast and resolution are considered for the sake of evaluation. More specifically, resolution is estimated as the full width at half maximum (FWHM) both in axial and lateral directions. The average value of FWHM among all scatterers in the image is reported. As for contrast, the contrast to noise ratio (CNR) is calculated as follows~\cite{7728908}:
\begin{equation} 
\label{eq:7}
\text{CNR} = 20\log_{10}(\frac{\mid \mu_{in}-\mu_{out}\mid  }{\sqrt{(\sigma_{in}^2+\sigma_{out}^2)/2}}),
\end{equation}
where $\mu$ and $\sigma$ are the mean gray level and the gray level standard deviation. Subscribes $._{in}$ and $._{out}$ refer to inside and outside
of the anechoic cystic region, respectively. Indexes are calculated on B-Mode images. In order to unify the comparison, we use the codes provided by PICMUS to compute the indexes. 
\subsection{Implementation Details}
\label{sec:sec33}
\begin{figure*}[b!]
	\centering
	\centerline{\includegraphics[width=0.8\textwidth]{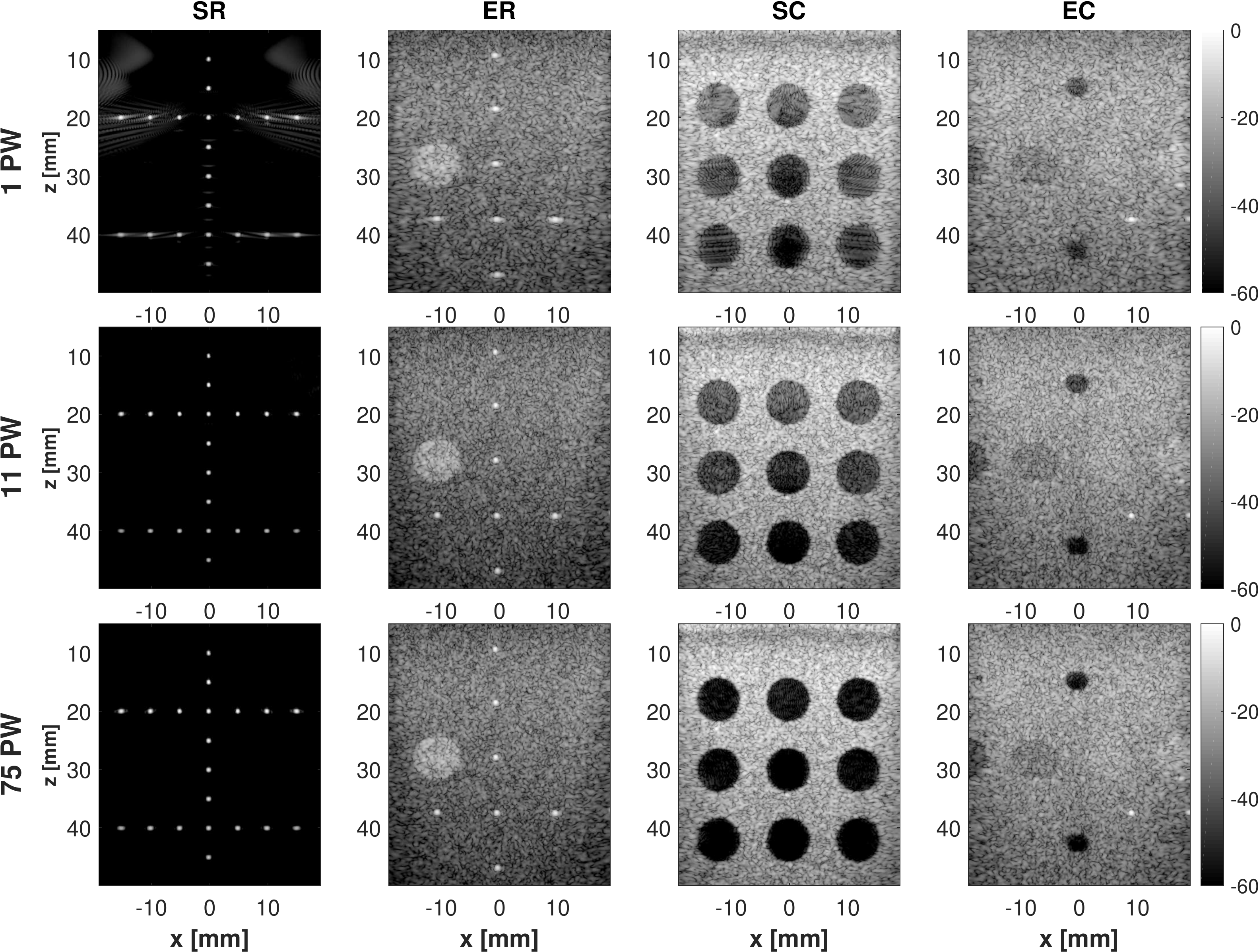}}
	\caption{ICA beamforming using 1, 11, and 75 plane waves. Columns indicate different image data sets and rows correspond to the number of transmitted plane waves.}
	\label{fig:fig3}
\end{figure*}
As explained in Section~\ref{sec:sec23}, the FastICA algorithm is used to estimate the apodization window. The maximum number of iterations is set to $100$ and the stopping criterion is set to be $\epsilon=10^{-6}$. The weights are initialized with random numbers extracted from standard distribution. The reduction of dimension through PCA is not used and the best results which are most reproducible are attained by considering  all eigenvalues in the estimation procedure. We use the Matlab implementation of the Fixed point ICA, the main algorithm of FastICA, which is publicly available online~\url{http://research.ics.aalto.fi/ica/fastica/code/dlcode.shtml}.\par
Throughout the results section, we consider $F=1.75$ and use Tukey (tapered cosine) window with constant parameter set to $0.25$ for DAS and other adaptive methods on top of DAS.\par
It is not possible to theoretically prove the convergence of FastICA algorithm with the mentioned parameters. In practice, however, we set the maximum number of iterations equal to $100$ and observe that for all of data sets, the algorithm converges in a lower number of iterations.
\begin{table}[t!]
	\caption{Quantitative results in terms of CNR and FWHM indexes for simulation and real phantom experiments.}
	\label{table:1}
	\centering
	\setlength{\tabcolsep}{2.5pt}
	\scriptsize
	\begin{tabular}{c c c c c c c c c c c c c c c c c}
		\specialrule{.15em}{0em}{.2em}
		dataset && SR & ER & SC & EC  \\ [.2em] 
		\specialrule{.05em}{0em}{.2em} 
		index && FWHM\textsubscript{A} FWHM\textsubscript{L} & FWHM\textsubscript{A} FWHM\textsubscript{L} & CNR & CNR \\ [.2em] 
		\specialrule{.05em}{0em}{.2em} 
		\makecell {1 PW} & \makecell{DAS \\ ICA} & \makecell{0.4 \,\,\,\,\,\,\,\,\,\,\ 0.82 \\0.39 \,\,\,\,\,\,\,\,\,\ 0.52} & \makecell{0.57 \,\,\,\,\,\,\,\,\,\ 0.88 \\0.57 \,\,\,\,\,\,\,\,\,\ 0.81}    & \makecell{\,\,9.95 \\\,\,10.67} & \makecell{8.15 \\8.9}  \\ [.2em] 
		\specialrule{.05em}{0em}{.2em}
		\makecell {11 PW} & \makecell{DAS \\ ICA} & \makecell{0.4 \,\,\,\,\,\,\,\,\,\,\ 0.54 \\0.4 \,\,\,\,\,\,\,\,\,\,\ 0.41} & \makecell{0.56 \,\,\,\,\,\,\,\,\,\ 0.54 \\0.56 \,\,\,\,\,\,\,\,\,\ 0.51} & \makecell{\,\,12.48 \\\,\,12.6} & \makecell{11.25 \\11.4} \\ 
		\specialrule{.1em}{0em}{.2em} 
		\makecell {75 PW} & \makecell{DAS \\ ICA} & \makecell{0.4 \,\,\,\,\,\,\,\,\,\,\ 0.56 \\0.4 \,\,\,\,\,\,\,\,\,\,\ 0.42} & \makecell{0.56 \,\,\,\,\,\,\,\,\,\ 0.56 \\0.56 \,\,\,\,\,\,\,\,\,\ 0.53} & \makecell{\,\,15.55 \\\,\,15.96} & \makecell{12 \\12.1} \\ 
		\specialrule{.1em}{0em}{.2em} 
	\end{tabular}
\end{table}
\section{RESULTS}
\label{sec:sec4} 
\subsection{Simulated and Experimental Data}
\label{sec:sec41}
The results of DAS beamforming versus our proposed method based on ICA on a single $0^{\circ}$ plane wave of simulated and experimental data are illustrated in Fig.~\ref{fig:fig2}. As seen from this figure, the proposed beamforming method outperforms DAS and improves the resolution as well as contrast on both simulated and experimental phantom data. In order to better investigate the amount of improvement, the quantitative indices are reported in Table~\ref{table:1}. What causes the improvement is the window used for apodization. So, as Table~\ref{table:1} confirms, improvement in resolution can only be acquired in the lateral direction. The highest improvement in lateral FWHM is $36.5\%$ on simulated plane-wave data of only one single transmission. In terms of CNR, approximately $10\%$ of improvement is achieved on the simulated cyst data of a single transmission while boarders of the cyst are also sharper.\par   
\begin{figure*}[b!]
	\centering
	\centerline{\includegraphics[width=\textwidth]{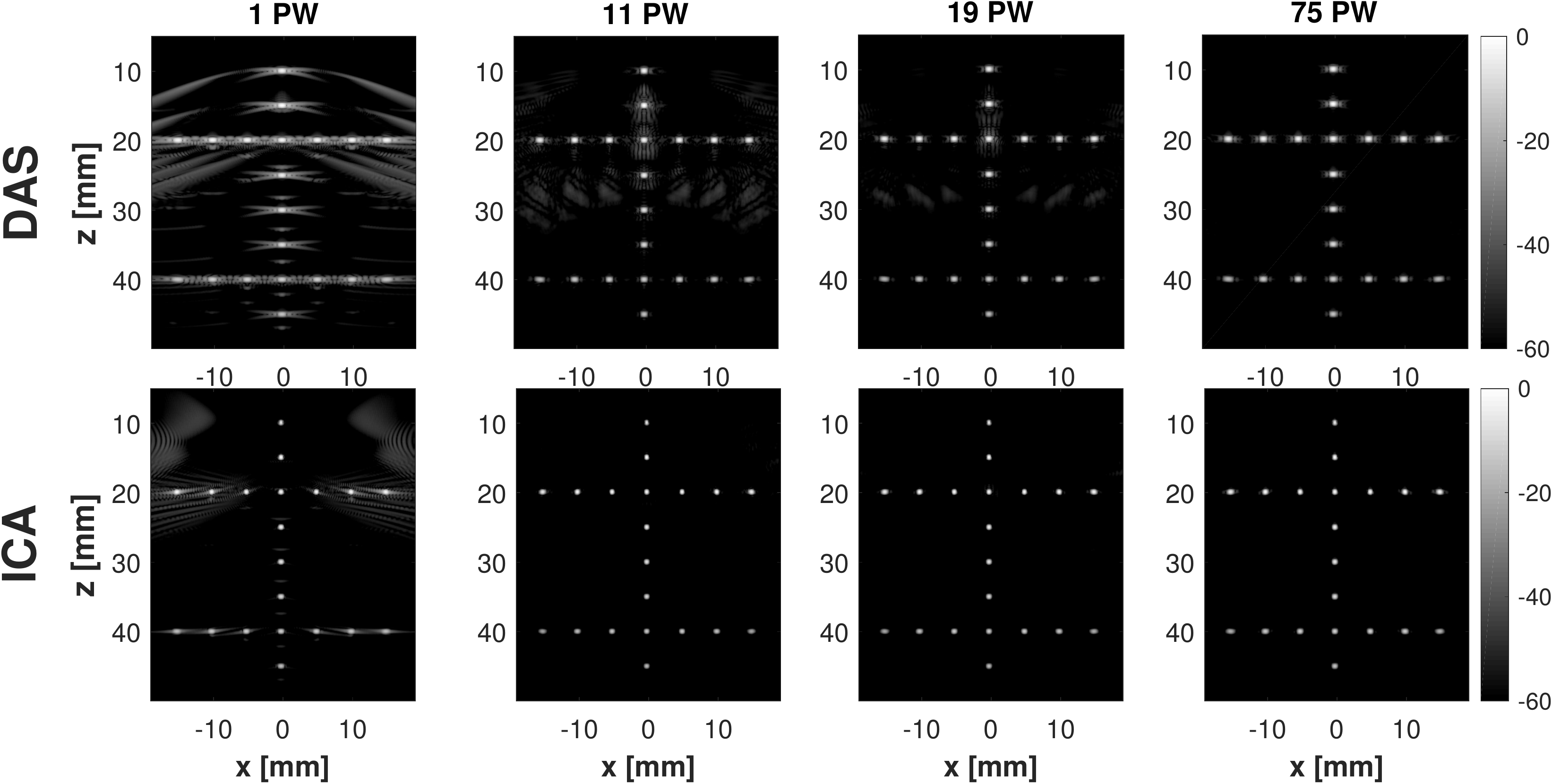}}
	\caption{Comparison of beamforming results using different number of plane waves. Rows indicate beamforming methods while columns correspond to the number of transmitted plane waves.}
	\label{fig:fig10}
\end{figure*}
In order to investigate the effect of CPWC, the results of the proposed method on higher number of plane waves are illustrated in Fig.~\ref{fig:fig3}. The indexes of Table~\ref{table:1} as well as Fig.~\ref{fig:fig3} confirm that CPWC improves the image quality.
As for CPWC, we do not repeat beamforming for each angle and use the apodization weights of $0^{\circ}$ plane wave for the remaining angles. Moreover, we do not apply any angular apodization weights to limit the sources of achieved improvement. In fact, our main focus is on beamforming of the received signals.\par
\begin{figure}[t!]
	\centering
	\centerline{\includegraphics[width=8.5cm]{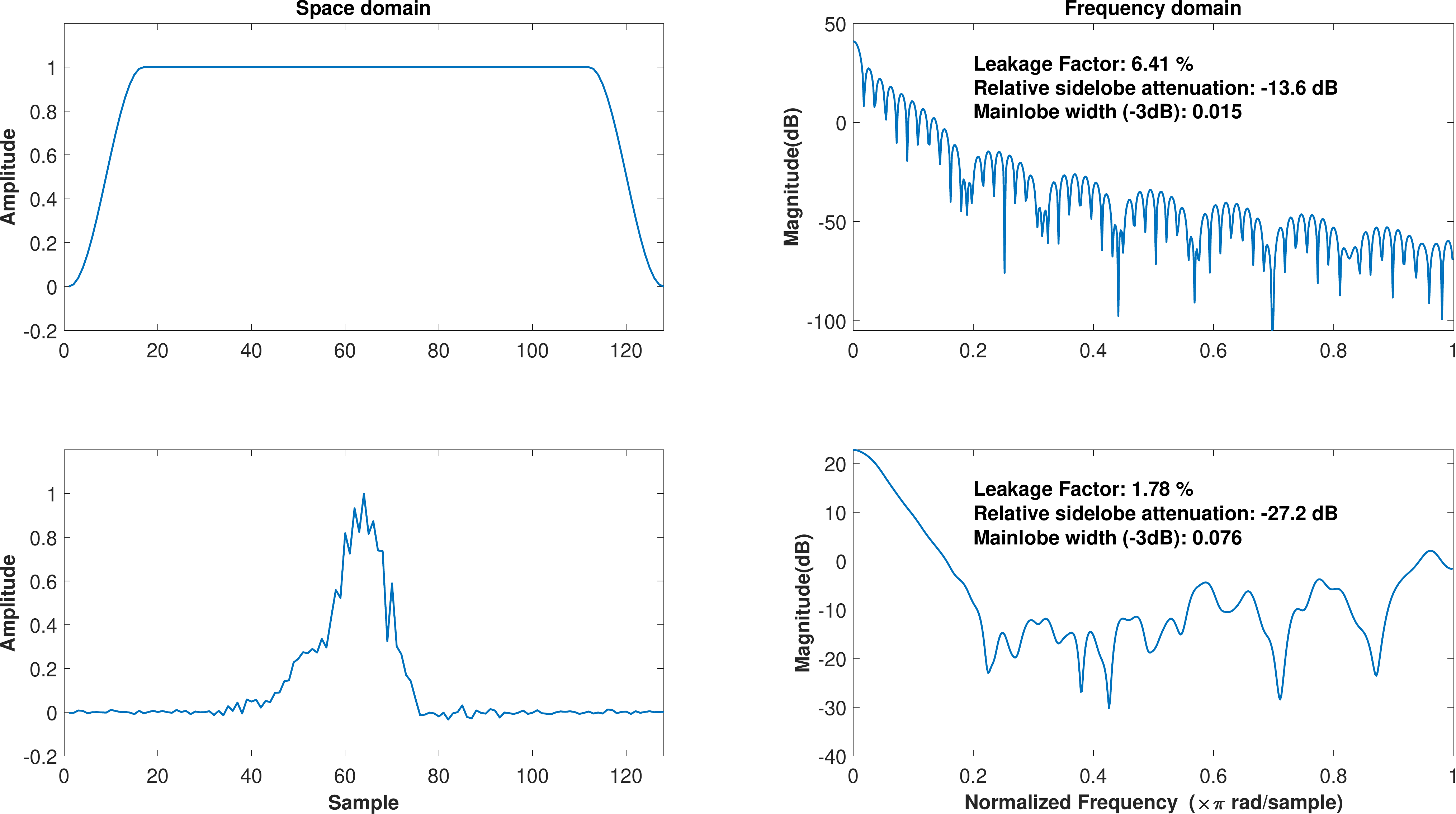}}
	\caption{Comparison of apodization window estimated using ICA (second row) and Tukey25 (first row) used in DAS. Windows are shown in both space and frequency domains.}
	\label{fig:fig4}
\end{figure}

 
To better understand the effect of proposed method, Fig.~\ref{fig:fig4} shows a comparison between Tukey25 window used in DAS and the apodization weights estimated by ICA on ER dataset. The estimated window has a lower leakage factor as well as a relative side lobe attenuation while its main lobe is wider. The estimated window is of a different shape which can not be found among predefined common windows. So, this point confirms the necessity of estimating the apodization window from the received data.\par
Fig.~\ref{fig:fig10} demonstrates qualitative improvements with ICA using fewer angles than needed with DAS. As can be seen in Fig.~\ref{fig:fig10}, the proposed approach achieves similar image quality with $11$ plane waves compared with DAS using $75$ angles. Therefore, it is possible to reduce the number of plane wave transmits needed to achieve image quality similar to a fully sampled transmit. 
\subsection{\textit{In vivo} Data}
\label{sec:sec42}
\begin{figure*}[t!]
	\centering
	\centerline{\includegraphics[width=0.8\textwidth]{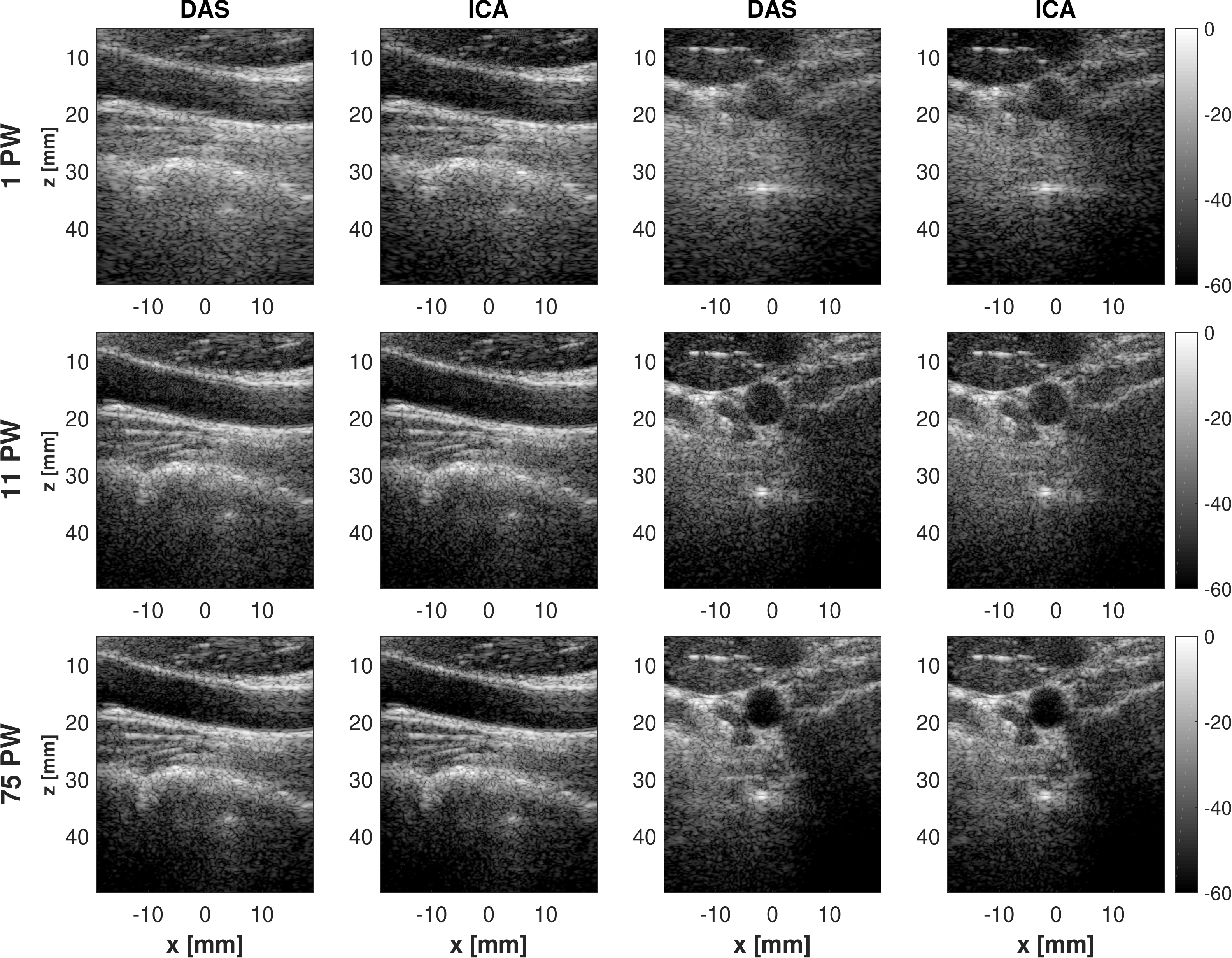}}
	\caption{Beamforming results on \textit{in vivo} data using 1, 11, and 75 plane waves. Two columns in left indicate cross-section images while left two columns correspond to longitudinal-section. Rows denote different number of transmitted plane waves used in beamforming.}
	\label{fig:fig5}
\end{figure*}

In real ultrasound tissues, there are more sources of degradation in image quality. In order to make sure that the proposed method also works on \textit{in vivo} data, the results of beamforming on real carotid images of PICMUS dataset are provided in Fig.~\ref{fig:fig5}. Visual comparison of beamformed images with different number of angles reveals that ICA outperforms classical DAS in both cross as well as longitudinal sections. Furthermore, ICA results in sharper images with a better contrast.\par  
\begin{figure}[b!]
	\centerline{\includegraphics[width=7.5cm]{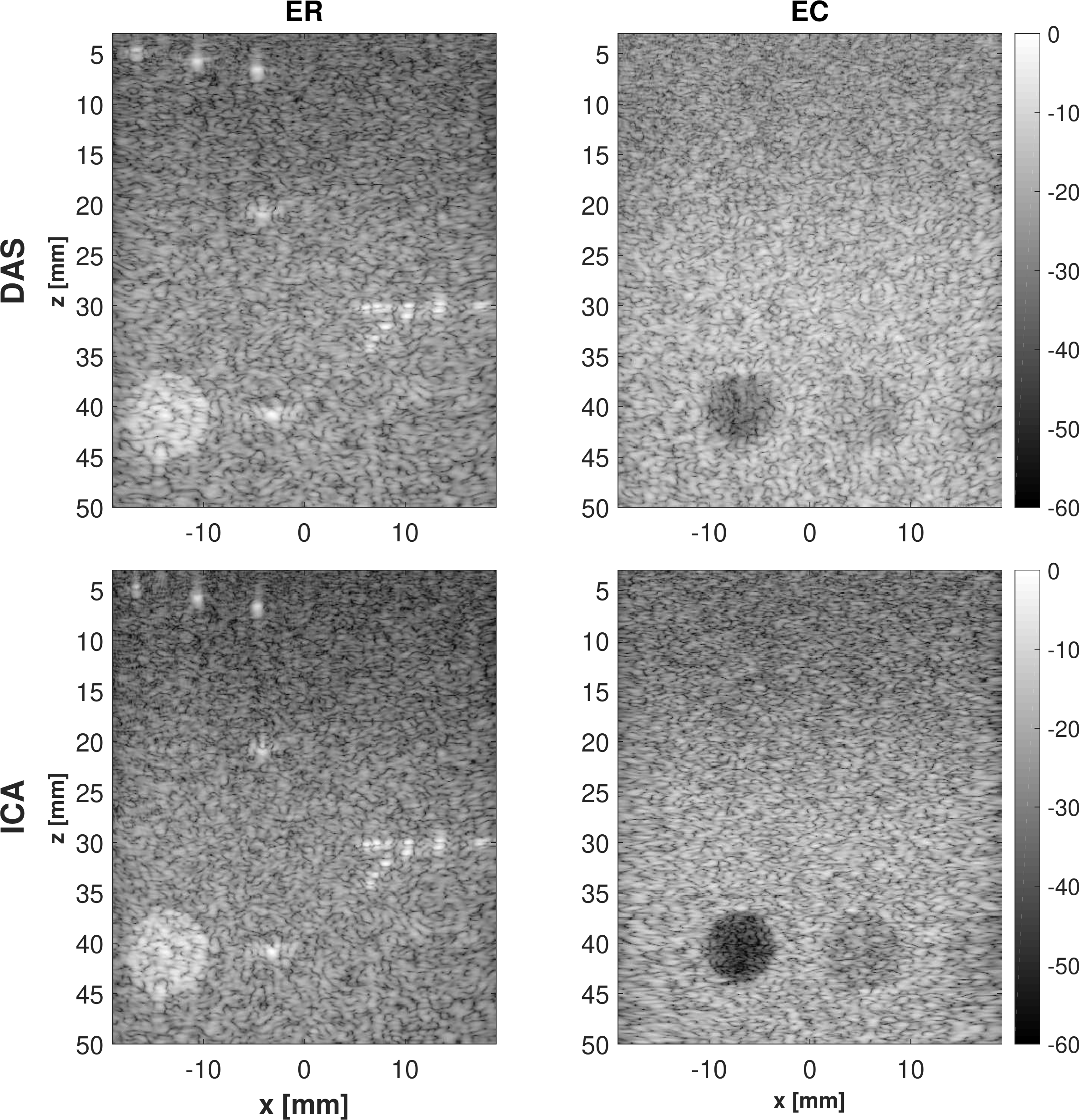}}
	\caption{Beamforming results on the single $0^{\circ}$ plane wave data collected with Alpinion Scanner. Columns indicate different image data sets while rows correspond to beamforming methods.}
	\label{fig:fig6}
\end{figure}
\begin{figure*}[t!]
	\centering
	\centerline{\includegraphics[width=0.8\textwidth]{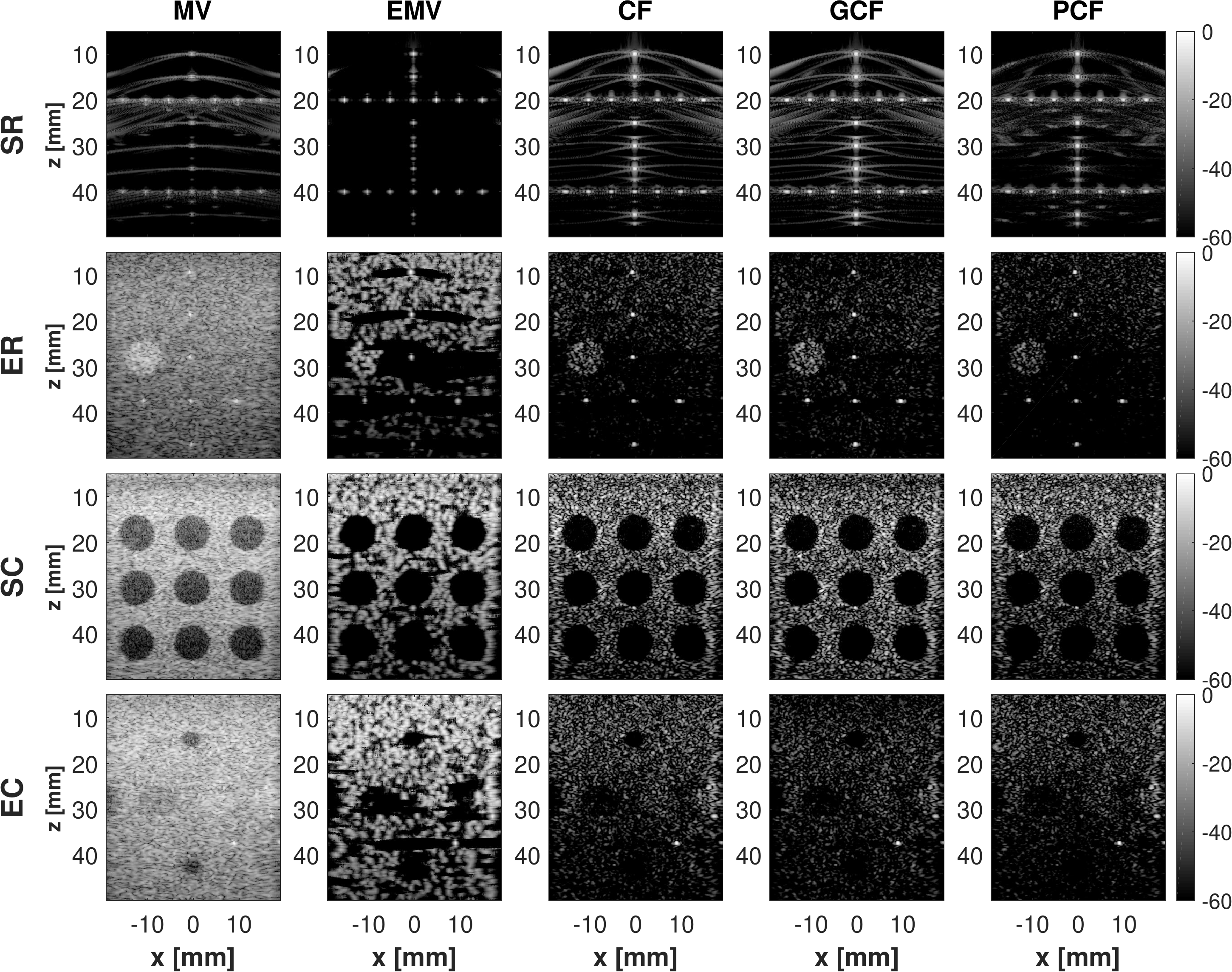}}
	\caption{Results of other adaptive beamforming methods on the single $0^{\circ}$ plane wave. Columns indicate different image data sets while rows correspond to different adaptive beamforming methods.}
	\label{fig:fig7}
\end{figure*}
\subsection{Results on Alpinion Scanner}
\label{sec:sec43}
We investigated the performance of our method on data collected with Alpinion Scanner thanks to ultrasound toolbox~\cite{8092389}. Fig.~\ref{fig:fig6} illustrates the result of proposed method on images of cyst regions and points scatterers recorded on an Alpinion scanner with a L3-8 probe from a CIRS phantom. The ICA results in Fig.~\ref{fig:fig6} obviously have a better quality as compared to DAS results in terms of resolution and contrast. As illustrated in Fig.~\ref{fig:fig6}, point targets in the image beamformed with ICA are finer in ER dataset, and contrast of cyst regions in EC dataset is noticeably improved. As we do not have the exact location of cysts and point targets in this dataset, quantitative comparison is not possible.\\
\\
\subsection{Comparison with Other Adaptive Methods}
\label{sec:sec44} 
As mentioned before, our focus in current study is on beamforming of the received signals. So, comparison with other adaptive approaches is of crucial importance. In this way, we present the results of five well-known approaches, namely MV~\cite{5278437}, EMV~\cite{5611687}, CF~\cite{1182117}, generalized CF (GCF)~\cite{1182117}, and PCF~\cite{4976281}. The comparison with these methods was not possible without using codes provided by Rindal \textit{et al.}~\cite{8691813} in ultrasound toolbox repository (~\url{http://www.ustb.no/publications/dynamic\_range/}). As for comparison with F-DMAS method, the sampling frequency of the PICMUS data was not large enough to apply the method with filtering around 2 times the center frequency. Fig.~\ref{fig:fig7} shows the result of different adaptive beamforming algorithms on a single $0^{\circ}$ plane wave of simulated and experimental data. The quantitative comparison is provided in Table~\ref{table:2}. The EMV method outperforms all other methods, even our proposed method, in terms of indices. However, methods based on the MV are very time consuming and are not practical for online applications. More specifically, the run time of our proposed method is in order of millisecond while EMV takes a few minutes. Furthermore, as it can be seen in Fig.~\ref{fig:fig7}, the EMV method destroys the image texture and its results are not visually appealing. Other approaches based on CF outperform the proposed approach in terms of FWHM index while are worse in terms of contrast.        
\subsection{Robustness to Noise}
\label{sec:sec45} 
In order to further demonstrate the superiority of our proposed method, the quality of beamforming approaches is investigated when some of the receiving channels are noisy. More specifically, we add independent Gaussian noise with different levels of signal to noise ratio (SNR) to different number of channels of ER dataset to see how noise affects the performance of different methods.\par
\begin{figure*}[t!]
	\centering
	\centerline{\includegraphics[width=0.7\textwidth]{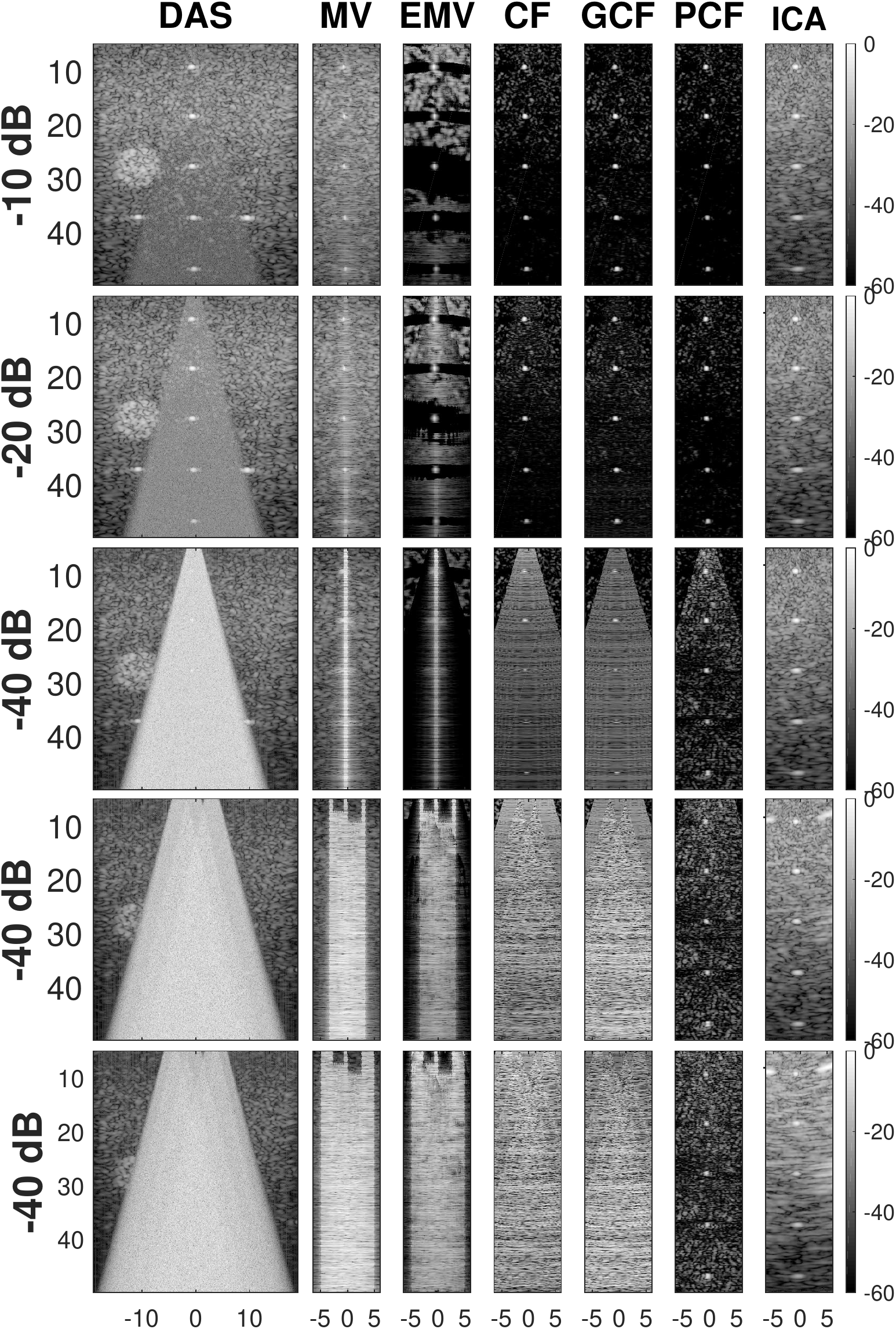}}
	\caption{Results of beamforming methods on the single $0^{\circ}$ plane wave of noisy ER dataset. Columns indicate different beamforming methods while three first rows correspond to different levels of noise added to channel $64$. Rows 4 and 5 correspondingly depict the results when we have 3 and 5 noisy channels with $-40dB$ SNR.}
	\label{fig:fig8}
\end{figure*}

The first three rows of Fig.~\ref{fig:fig8} illustrate the results for additive Gaussian noise with $-10$, $-20$, and $-40dB$ SNR levels with noise added to channel $64$ (the total number of channels is $128$). As seen in Fig.~\ref{fig:fig8}, other adaptive methods and DAS fail to reconstruct the affected region with noise while the proposed beamforming algorithm is completely robust. The fourth and fifth rows of Fig.~\ref{fig:fig8} correspondingly show the results when $-40dB$ additive noise is added to 3 and 5 of crystal elements. As the number of noisy channels with a bad SNR goes up, the quality of ICA starts to decrease. However, it still outperforms other approaches.    
\begin{table}[b!]
	\caption{Quantitative results of other adaptive beamforming methods in terms of CNR and FWHM indexes for simulation and real phantom experiments.}
	\label{table:2}
	\centering
	\setlength{\tabcolsep}{2.5pt}
	\scriptsize
	\begin{tabular}{c c c c c c c c c c c c c c c c c}
		\specialrule{.15em}{0em}{.2em}
		dataset && SR & ER & SC & EC  \\ [.2em] 
		\specialrule{.05em}{0em}{.2em} 
		index && FWHM\textsubscript{A} FWHM\textsubscript{L} & FWHM\textsubscript{A} FWHM\textsubscript{L} & CNR & CNR \\ [.2em] 
		\specialrule{.05em}{0em}{.2em} 
		\makecell {1 PW} & \makecell{MV \\ EMV \\ CF \\ GCF \\ PCF} & \makecell{0.41\, \,\,\,\,\,\,\,\,\,\,\ 0.1 \\0.39 \,\,\,\,\,\,\,\,\,\ 0.09\\0.32 \,\,\,\,\,\,\,\,\,\ 0.44\\0.32 \,\,\,\,\,\,\,\,\,\ 0.43\\0.29 \,\,\,\,\,\,\,\,\,\ 0.37} & \makecell{0.59 \,\,\,\,\,\,\,\,\,\ 0.43 \\0.58 \,\,\,\,\,\,\,\,\,\ 0.33\\0.48 \,\,\,\,\,\,\,\,\,\ 0.47\\0.48 \,\,\,\,\,\,\,\,\,\ 0.47\\0.46 \,\,\,\,\,\,\,\,\,\ 0.41}    & \makecell{\,\,11.1 \\\,\,14\\\,\,8.2\\\,\,8.1\\\,\,6.9} & \makecell{7.95 \\10.5\\6.3\\6.3\\5.2}  \\ [.2em] 
		\specialrule{.05em}{0em}{.2em}
		\makecell {11 PW} & \makecell{MV \\ EMV \\ CF \\ GCF \\ PCF} & \makecell{0.43\, \,\,\,\,\,\,\,\,\,\,\ 0.1 \\0.4 \,\,\,\,\,\,\,\,\,\,\ 0.09\\0.37\, \,\,\,\,\,\,\,\,\,\,\ 0.37\\0.38 \,\,\,\,\,\,\,\,\,\,\ 0.36\\0.37 \,\,\,\,\,\,\,\,\,\,\ 0.3} & \makecell{0.59 \,\,\,\,\,\,\,\,\,\ 0.29 \\0.56 \,\,\,\,\,\,\,\,\,\ 0.28\\0.55 \,\,\,\,\,\,\,\,\,\ 0.37 \\0.55 \,\,\,\,\,\,\,\,\,\ 0.37 \\0.55 \,\,\,\,\,\,\,\,\,\ 0.31} & \makecell{\,\,11.4 \\\,\,15.2\\\,\,11.9\\\,\,11.8\\\,\,11} & \makecell{9.8 \\11.5\\10.2\\10.2\\9.05} \\ 
		\specialrule{.1em}{0em}{.2em} 
		\makecell {75 PW} & \makecell{MV \\ EMV \\ CF \\ GCF \\ PCF} & \makecell{0.43 \,\,\,\,\,\,\,\,\,\,\ 0.1 \\0.4 \,\,\,\,\,\,\,\,\,\,\ 0.09\\0.4 \,\,\,\,\,\,\,\,\,\,\ 0.38\\0.4 \,\,\,\,\,\,\,\,\,\,\ 0.38\\0.39 \,\,\,\,\,\,\,\,\,\,\ 0.29} & \makecell{0.58 \,\,\,\,\,\,\,\,\,\ 0.31 \\0.56 \,\,\,\,\,\,\,\,\,\ 0.29 \\0.56 \,\,\,\,\,\,\,\,\,\ 0.38 \\0.56 \,\,\,\,\,\,\,\,\,\ 0.38 \\0.56 \,\,\,\,\,\,\,\,\,\ 0.32} & \makecell{\,\,14.7 \\\,\,17\\\,\,14.05\\\,\,13.9\\\,\,14.13} & \makecell{11 \\10.4\\10\\10\\10.3} \\ 
		\specialrule{.1em}{0em}{.2em} 
	\end{tabular}
\end{table}
\section{DISCUSSION}                  
Using a part of samples of each channel which only correspond to the middle part of the final image is important from two aspects. First, it removes the effect of incomplete data of borders on the ICA performance. Second, the FastICA algorithm converges faster as it works with a lower amount of data. Note that the estimated apodization weights and its specifications such as width of main lobe or the amount of side lobe attenuation in each dataset are different. So, it can be concluded that there is not a unique solution that works for all data.\par
The algorithm can be applied for each angle separately. However, the estimated weights for different insonification angles are not totally different and the improvement is negligible while the processing time is multiplied corresponding to the number of angles. The angular apodization can also be performed using ICA for CPWC. However, the main focus of this study was apodization of the received signals. The angular weights are not used to limit the sources of improvement, which make the comparison with other approaches possible.\par
As each adaptive beamforming method alters the dynamic range of the final image differently, we normalized the range of final outputs to plot the results in a uniform manner and make the visual comparison possible.\par
Robustness of the proposed method to noise can be explained as the ability of ICA in extracting independent components and discarding noise. This aspect of the proposed beamforming algorithm is very important in practice when we have a missing or noisy channel. Although a malfunctioning element affects both transmit and receive, the available data does not give us the chance of a perfect simulation of this case. In future, we plan to test the
performance of the proposed method for the case of having broken channels in both transmit and receive.    
\section{CONCLUSIONS}
We have proposed a new beamforming approach for ultrasound plane-wave imaging based on ICA. Beamforming has been formalized as the estimation of one independent image out of several non-independent observation and the apodization weights have been estimated based on collected data. The images of one single plane-wave transmission as well as multiangle plane-wave acquisitions have been successfully reconstructed. Moreover, the performance of the algorithm has been demonstrated on different imaging settings. Results show that the proposed method simultaneously improves the resolution and contrast while the resulting image is also visually appealing. Furthermore, the proposed algorithm is strongly robust when some of channels are noisy.   
\section*{Acknowledgement}
This project was funded by NSERC Discovery grants RGPIN-04136 and RGPIN2017-06629. The authors would like to thank the organizers of the PICMUS challenge as well as the ultrasound toolbox for providing publicly available codes and data. 
\appendices
%
\bibliographystyle{IEEEbib}
\bibliography{refs}
\end{document}